\documentclass{ae}[times]  

\usepackage{graphicx}
\usepackage{txfonts}
\usepackage{hyperref}
\begin{document} 

   \title{{\em \LARGE{Very Important\\Letter to the Editor}}\vspace{0.6cm}\\ On the Use of Astronomy: II. The secret of the elixir of youth of blue straggler stars}
\titlerunning{The secret of the elixir of youth}
\authorrunning{Henri et al.}

   \author{Henri M.J. Boffin
          \inst{1}\fnmsep\thanks{As for all his papers, this author is sole responsible for its content, which does not represent in any way or another, not even when seen through a telescope, the views of his employer, real or supposed.}
          \and
          A. Wake\inst{2}
          \and
          W.H.Y. Can't\inst{2}, 
          I. Sleep\inst{2}
          }

   \institute{Extraterrestrial Institute for the Advancement of Earth (EIAE),
            Secret place, Planet Earth, Solar System  
                 \and
             The Improbable Institute, Flatland
             }

   \date{Received March 30, 2021; accepted March 31, 2021}

 
  \abstract
   {Using Gaia EDR3, we study the most spectacular and photogenic cluster of Ptolemy.  After deriving its membership, we identify in its colour-magnitude diagram a star that definitively decided to straggle and dress in blue. Further analysis with the FARCE telescope allows us to discover in its light curve the secret of its rejuvenation, which we gladly share in this paper. This research is an important contribution to attain the ultimate goal of astronomy as professed by DJ Format.
   \vspace{0.5cm}}
   \maketitle
\begin{flushright} {\it We are all in the gutter, but some of us are looking at the stars}\\  -- Oscar Wilde
\end{flushright}

\section{The ultimate goal of astronomy}
\begin{figure}
    \begin{center}
        \includegraphics[width=9cm]{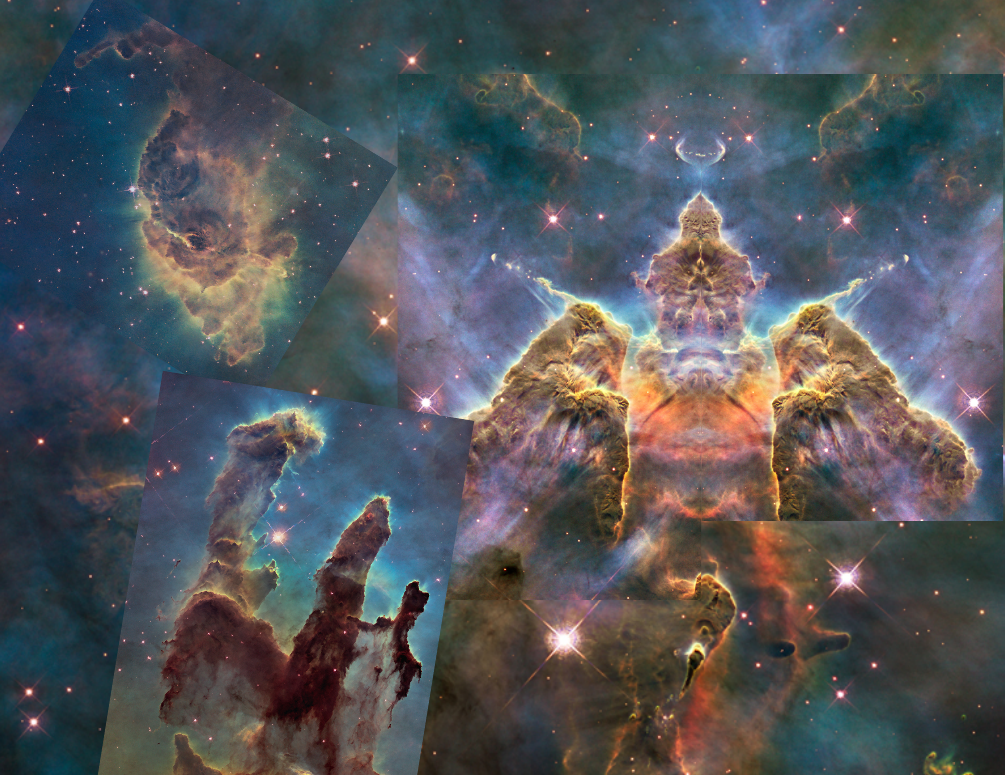}
    \caption{Collage of Hubble Space Telescope photographs of some parts of the Carina Nebula. Although this image has nothing to do with the science presented in this paper, it is, we think, a nice illustration that astronomy can also be artistic. { \tiny Credit of the individual images: NASA, ESA, the Hubble SM4 ERO Team, M. Livio, the Hubble 20th Anniversary Team (STScI), and the Hubble Heritage Team.}
    }
    \label{fig:art}
    	\end{center}
\end{figure}

According to \cite{dj}, who is without any doubt a disciple of Jeremy Bentham, ``{\it starting a colony deep in the galaxy must be the aim of astronomy}'', which is a standpoint we can't agree with\footnote{But we can't help agreeing with him when he tells us to ``{\it  notice the atom's autonomy physically mimicking that of the planets in gravity.}''}, as we are convinced that astronomy can serve other purposes, all very noble. For example, there is no doubt in our minds that astronomy possesses a clear artistic dimension, a proof of which is given in Fig.~\ref{fig:art}. Thus, 
the authors of this paper, in their combined wisdom and with the hope that the referee and the editor are not unconditional fan of DJ Format and would thereby look askance at this contribution, intended to exert their free will and address what they consider to be yet another essential contribution to astronomy, and one not quixotic at all. This, in hindsight, turned out to be -- once again!-- most illusory, as the important result that is presented in this dithyramb, which is, shall we dare say, not only original but bound to become seminal, is nothing else than the stepping stone needed to achieve the aforesaid goal. 

Our intention here is to look indeed at blue straggler stars, and this is no bathos, we assure you. Society, and likely your parents as well, warn against the effects of straggling, and more than a gardener will cut with a sadistic pleasure these branches straggling from their bush. But, in astronomy, it is often the case that what shouldn't exist proves the most interesting. And indeed so are blue straggler stars, when they are not of the Hollywood kind. Stars, these natural thermonuclear reactors that shine and often twinkle, twinkle when they are little, have a given lifetime, which, not very unlike most representatives of the Humankind, depends on their mass. The more massive a star, the less it will live. Still, some facetious celestial bodies seem to outperform their life expectancy. They thus appear to have found an elixir of youth, looking younger than they really are. As this is a dream for many people on planet Earth, it therefore seems that, once more, astronomy may prove most useful: if we manage to find out the secret of why some stars appear younger than they are, this could have many practical applications. One of them being to make it easier for people to travel to outer space, as they wouldn't age so quickly, and thus making DJ Format happy. This is thus an excellent subject for the third paper in this series, after having shown that it is quite likely that all non-avian mammals will disappear 12,194,755 years from now \citep{bof14} and that there exists a stellar eroteme in the sky, which is Nature's way of telling us, in the most beautiful way, that we still don't understand planetary nebulae \citep{bof20}. 

It is important to note that, although we suppose that our three courageous readers, who are no churls, are all aware of it, let us stress here that to the best of our knowledge, we have exerted the utmost care not to use any anacoluthon, anaphora, asyndeton, aposiopesis, hyperbaton, nor tmesis in this portentous paper, so as to make its reading easier, and, hopefully, even sybaritic.  

\section{The open bunch}
\begin{figure}
    \begin{center}
        \includegraphics[width=9cm]{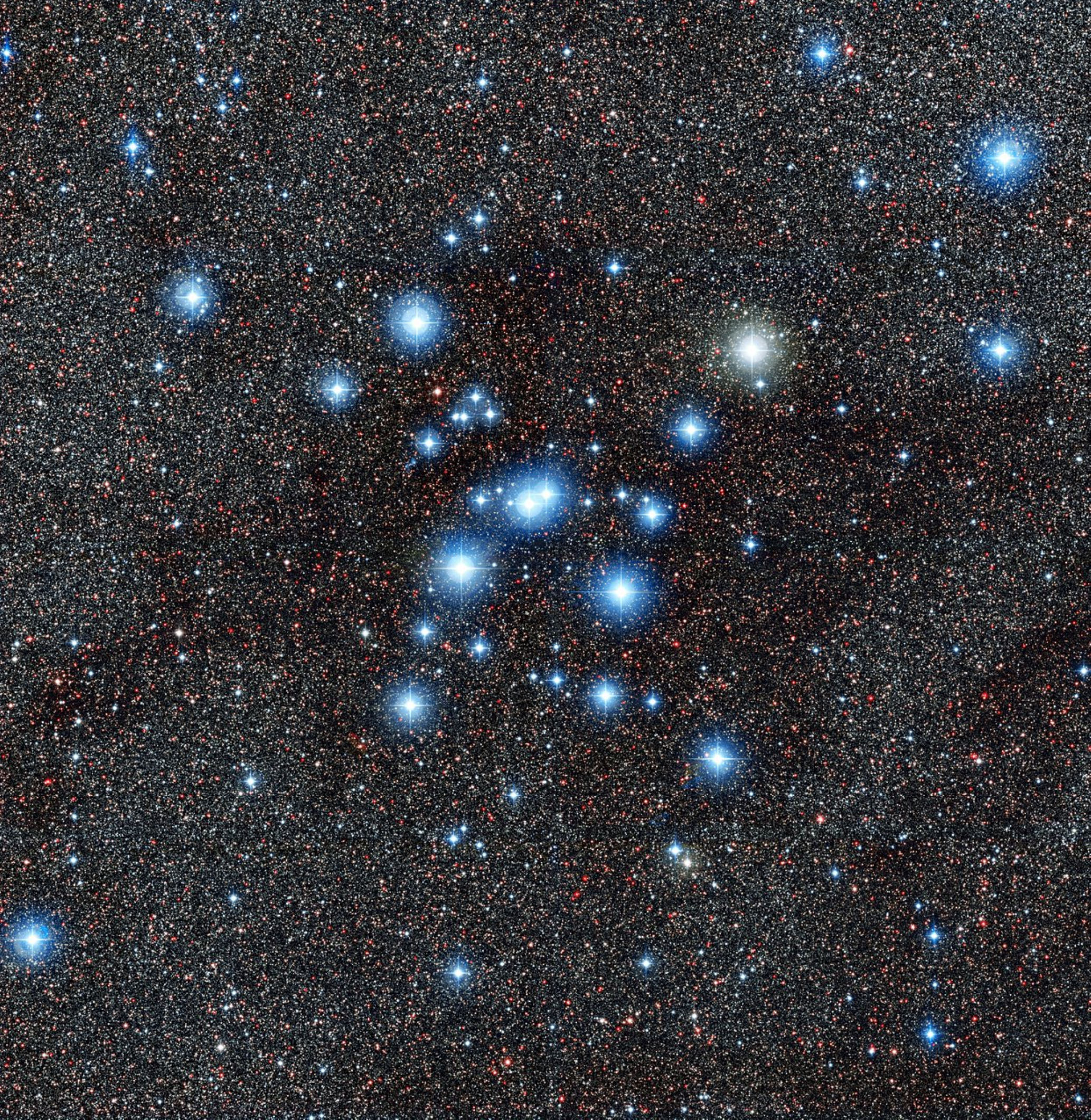}
    \caption{This image from the Wide Field Imager on the MPG/ESO 2.2-metre telescope at ESO's La Silla Observatory in Chile shows the central parts of bright star cluster Messier 7, also known as NGC 6475. Easily spotted by the unaided eye in the direction of the tail of the constellation of Scorpius (The Scorpion), this cluster is one of the most prominent open clusters of stars in the sky and an important research target.
{ \tiny Credit: ESO.}}
    \label{fig:M7}
    	\end{center}
\end{figure}

In the same way that Humans prefer -- that is, until most recently -- live in cities, stars have a tendency to form in aggregates or clusters, which can be globular or open. As we are in favour of the greatest amount of freedom, and a fierce guardian of open source, whether in science or in software, we will consider here only open clusters. These are groups of up to a few thousand stars that formed together and thus share the same age and same chemical composition, just as millennials do among Earth's population. This is particularly useful for astronomers as they can study in detail a given population of stars. Nice examples of such open clusters are the Hyades or the Pleiades, but as our budget did not allow us to study these celebrities, we will look here at the open cluster Messier 7, also known as Ptolemy's Cluster, as it was first recorded by the astronomer Ptolemy, who, according to Wikipedia, mistakenly described it as a nebula in 130 AD. A glance at Fig.~\ref{fig:M7} will, however, convince even the most demanding of our readers that this is a very admirable specimen that deserves care and attention. 

The cluster, which is about 200 million year old -- that is, basically still a toddler as far as clusters go -- is located 280 pc away. However, to go beyond the mere pleasing of the eyes, even if very rewarding (and move towards a pleasing of the mind), one needs to establish a nose count of the cluster -- in as much as stars have noses. This requires winnowing real members in the warrens of the clusters from field contaminants that are there just to crash the party and get free food from the buffet. Our savvy readers will no doubt assert with the bold assurance that characterises them that to establish if a star is a member or not of a cluster, is just a matter of knowing if they are paying their membership fee. Unfortunately, the Universe bears in this respect an unfortunate likeness to the most select clubs in that they do not want to share their list of members. In fact, it is most likely that the membership book of the Universe is hidden behind a door that is guarded by the terrifying monster shown in Fig. \ref{fig:art}. We are not of the opinion asserted by some eristics that ``brave astronomer'' is an oxymoron, as we know some who are most courageous and even daredevils -- how else would you explain that they don't hesitate to embark in the most daunting task of peering inside black holes? -- but it is certain also that astronomers aren't fools and know their limits. They prefer thus leaving the behemoth to play with his cosmic dog and teapot\footnote{Is this really what you see in this image? -- Ed.} and find other ways to establish membership. 

\section{Cooking like an astronomer}
Thus, astronomers had to go to the great burden of putting in space a dedicated spy satellite, Gaia \citep{2016A&A...595A...1G}, whose unique goal is to measure precisely where the stellar members of the Milky Way are located and how they move. This can then be used to assert membership to a given cluster. In this paper, we have used data from the early third data release\footnote{\url{https://gea.esac.esa.int/archive/}} 
\citep{2020arXiv201201533G}. To perform such selection is an art more than a science and like good cooks, every astronomer has their own recipe on establishing memberships. Some even decide to complement the likely great power of their brain with artificial intelligence -- a procedure that of course leads to a direct disqualification from the Olympic Games of astronomy, but does advance science sometimes. And like good cooks, we will not bother explaining to you here our own recipe, as it is not the steps that count, but the {\it savoir-faire}! We will just mention here that as the authors of this work share, in addition to a liking for all beautiful things -- except when they are poisonous --, an utter aversion for these limbless scaly elongate reptiles called snakes, among which the {\it Pythonidae} family is no exception, they preferred to use for their analysis, as a fond recollection of the innocent times of their childhood when they thought that everything was possible, the ABC programming language\footnote{\url{https://homepages.cwi.nl/~steven/abc/}}. It also felt most appropriate to use this Dutch-made alphabet as a wrapper to the GPU-Optimised  Underperforming  Device  Architecture (GOUDA), that was put into service to achieve our goals. The tip of the iceberg of the wonderful achievement of this investigation is  shown in Fig.~\ref{fig:cmd}, where the impressive main sequence in the colour-magnitude diagramme is clearly seen. There is no word to describe the emotion that overtook the present authors when they first witnessed the outcome of their recipe from the GOUDA. The diagramme not only shows particularly well the main-sequence along which stars of a given mass have a well defined colour ($B_p-R_p$, which is a proxy for the stellar surface temperature) and brightness ($G$), but also some outliers that cannot hide the fact that they are actually distichs\footnote{Called binary stars by the majority of astronomers. We prefer this denomination.}: given the contribution from a companion, the system is shifted in the diagram, as it is hotter and more luminous than a single star. Our convolved, Byzantine, and tedious (but never tortuous) analysis allowed us to infer a cluster's membership of 591 stars -- a dramatic increase from the initial investigation done by Italian astronomer Giovanni Battista Hodierna around 1654 and who counted 30 stars in it! This information has immediately been forwarded to the tax office of the Universe so that they can update their files, but this is not the most important here. 
	
Before highlighting the most important, we need to mention here a caveat of our methodology. The data provided by Gaia are unfortunately missing one parameter. This reminds the first author of this noteworthy contribution, of a personal experience, which is as painful as inevitable: whenever he travels to some far place, he always, but always, in a way that is almost obsessive, forgets one thing. It is always a different thing, but nevertheless, every time he forgets to bring something. Here with Gaia, we are missing the radial velocity, that is, the speed at which the star is moving towards or away from us. What we know of our stars is where they are, how far away they are and how they move in the plane of the sky. But whether the star is moving at a blazing speed towards us, such that it will soon create havoc in our Solar System, or is in fact going away as if it couldn't care less about us, is something we unfortunately do not know. The perfect alignment of stars seen in Fig.~\ref{fig:cmd} indicates, however, that this is not of great concern for our study and we will therefore ignore this very annoying feature.

\section{The happy blue straggler}
\begin{figure}
    \begin{center}
        \includegraphics[width=9cm]{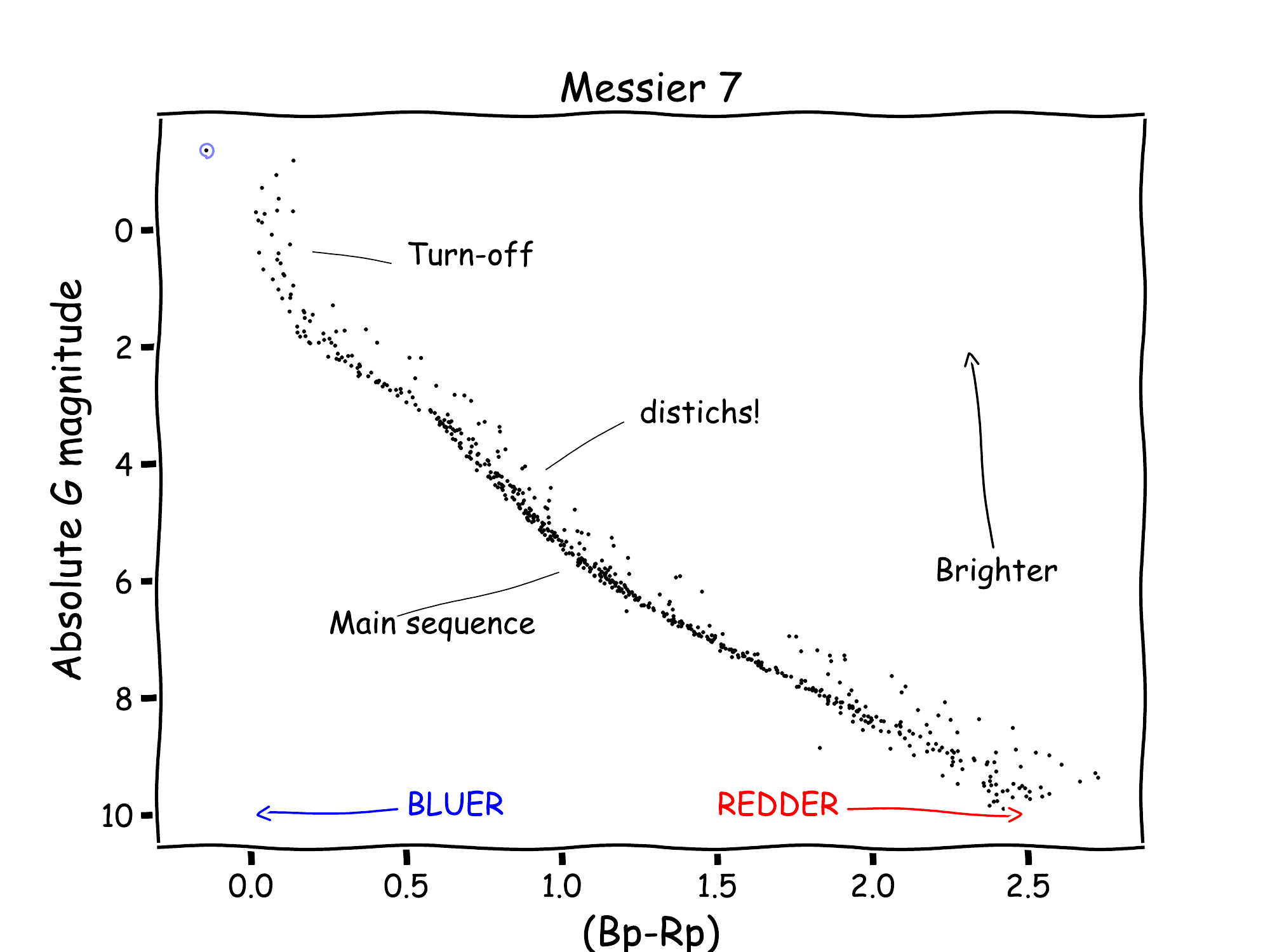}
    \caption{Colour-magnitude diagramme of Ptolemy's Cluster, based on Gaia-determined membership. The star surrounded by a blue circle is the happy blue straggler star.}
    \label{fig:cmd}
    	\end{center}
\end{figure}

The most careful examination of Fig.~\ref{fig:cmd} reveals, however, the presence of an asocial member. The main sequence of stars in the colour-magnitude diagramme presents the typical shape of a damaged hockey stick, as above a given mass, the time spent on the main sequence (that is, the time when they are frantically burning hydrogen in their core without fear of tomorrow) is smaller than the age of the cluster: the star would thus turn right -- this is the turn-off! -- to become a red giant and finally fall into oblivion as a white dwarf. It is well known that in all parties, if most are happily following the required pattern and fit into the mould, there is always a guy that has decided to go on another path. This is also the case here: the star whose symbol is surrounded by a blue circle in Fig.~\ref{fig:cmd} is located in a part of the diagramme where it shouldn't be -- it has straggled towards the blue! It is indeed too massive to still be on the main sequence and should have long since become a red giant. But, for some reason that we will try to find out, it managed to become young again, younger than the age of the cluster!

In this particular case, this blue straggler star is entry number 6647 of the bright star catalogue (HR 6647). The Gaia EDR3 tells us, with an amazing level of detail that you shouldn't trust, that its parallax is 
$3.6061993706266136 \pm 0.06444364$ milli-arcseconds, and that it is thus located $277.300253598081 \pm 4.955575241028729$ parcsecs\footnote{Students of the world, unite in reason: please do not follow the Gaia example when quoting values with their errors and only provide the significant numbers. Here, the object is located $277 \pm 5$ pc away, that's all we can say!} away -- it is therefore, for what we are concerned with, at the same distance as the cluster. Because this straggler is so bright, we already know a little bit about it: it has a surface temperature of between 15,500--17,000 K; it is about a thousand times more luminous than the Sun; and it has likely a mass of $5.21 \pm 0.27$ solar masses  \citep{1982A&A...109...37M,2012A&A...537A.120Z,2014A&A...566A.132S}. The same authors found out that the star has an apparent age equal to 0.785 times the age of the cluster. Now, this is a rejuvenated star! 

\section{Finding the secret}
The authors of this contribution, whose essential nature is, we are sure, becoming more and more obvious to the readers, have stopped at nothing to unravel the mystery and like Prometheus, bring its secret to the people. To this aim, the first author managed to rescue from the bin, where his wife had thrown it after his son decided this was no more for him, the FARCE  instrument \citep{joke}  attached to the 5-cm Extremely LIttle TElescope  \citep[ELITE; ][]{bof14,bof20}. This was then used to measure the flux in the $W$-band of the star as a function of time, taking one image every minute, for the whole visibility period of the star. The resulting light curve presents essentially random noise and is therefore not shown here in order to save the planet by shortening this paper as much as possible.

\begin{figure}
    \begin{center}
        \includegraphics[width=9cm]{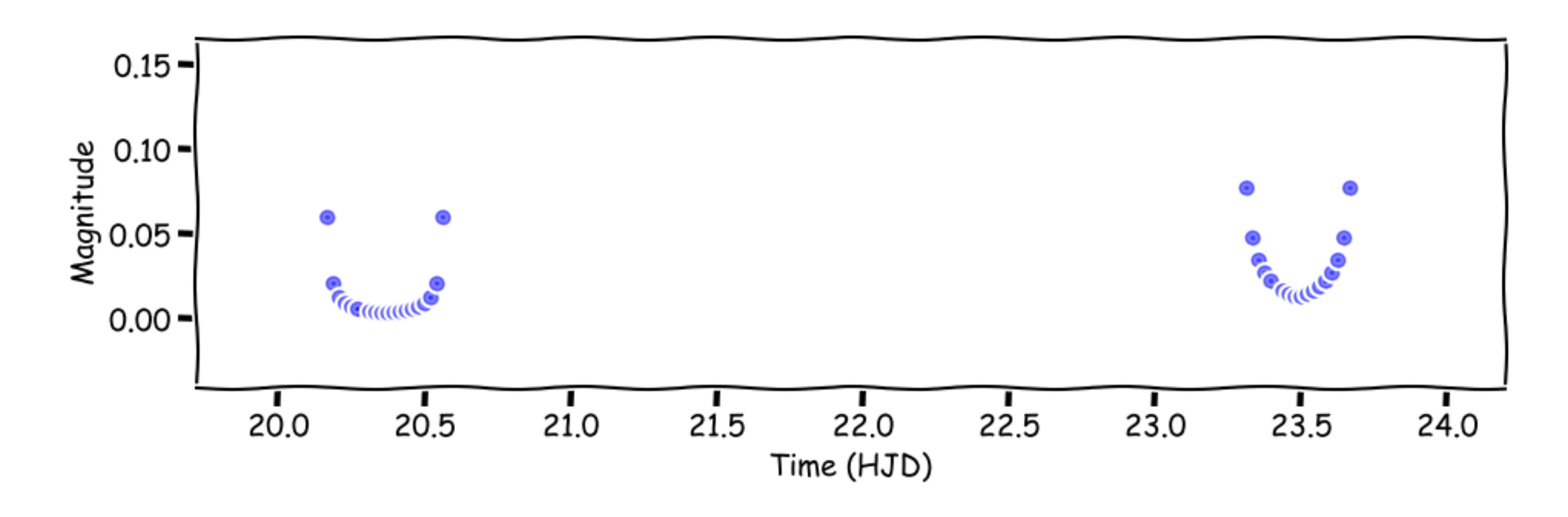}
    \caption{The secret of the blue straggler star as revealed by its light curve!}
    \label{fig:smile}
    	\end{center}
\end{figure}

Because the first author is the proud father of two, he isn't scared nor annoyed anymore by noise and so looked even deeper in the data to see if he could make any sense of it. The outcome of this daunting task is likely the lifelong achievement of the present authors. It is a revelation that will mark mankind for the next millennia. A close look at some parts of the light curve reveal indeed the secret of the elixir of youth that this blue straggler star must have drunk (or fallen into when it was a kid!). As seen in Fig.~\ref{fig:smile}, the star is showing at regular intervals a smiley! This is thus the secret of becoming younger: make sure to regularly smile.
We hope that by reading this paper, we managed to influence your brain to trigger on several occasions a reaction in which the zygomatic muscles in your cheeks have contracted and lift the corners of your lips, and thereby made you a blue straggler, which became slightly younger than it used to be.

\begin{acknowledgements}
As for the two other master pieces in this series, this work was done outside of working hours, when the first author was bothered by his co-authors. 
And here the obligatory stuff: this work has made use of data from the European Space Agency (ESA) mission
{\it Gaia} (\url{https://www.cosmos.esa.int/gaia}), processed by the {\it Gaia}
Data Processing and Analysis Consortium (DPAC,
\url{https://www.cosmos.esa.int/web/gaia/dpac/consortium}). Funding for the DPAC
has been provided by national institutions, in particular the institutions
participating in the {\it Gaia} Multilateral Agreement.
Matplotlib \citep{2007CSE.....9...90H} and its xkcd environment are gratefully acknowledged.
\end{acknowledgements}

\end{document}